# Building Europe’s Quantum Shield: The Strategic view for a Continent-Wide Quantum Key Distribution (QKD) Infrastructure

**Leandros Maglaras[1], Ilias Papastamatiou[2], Alexios Aivaliotis[2], Evangelos Markatos[3], Kontantinos Karantzalos[1]**

[1]*Ministry of Digital Governance, 11 Frangoudi & Pantou Avenue, Kallithea, Greece*
[2] *National Infrastructures for Research and Technology , 7 Kifisias Avenue, Athens, Greece*
[3]*Foundation for Research and Technology, Distributed Computing Systems and Cybersecurity, Nikolaou Plastira 100, Vassilika Vouton, Heraklion, Greece.*

*Abstract:* The fast growth of quantum computing can lead to amazing scientific breakthroughs while on the same time can be used to break today's security systems, raising new risks for existing digital systems. Facing this challenge, the European's Union's deployement of the European Communication Infrastructure (EuroQCI) is crucial. The SEEWQCI project combines fiber cables, satellite communications and enhanced security rules to build a strong digital shield. Its focus is to protect vital services like power grids and hospitals keeping Europeans' data safe..

## 1. Introduction

The global digital ecosystem stands on the edge of a crucial technological shift driven by quantum physics. Quantum computing, which uses the principles of superposition, entanglement, and quantum interference, promises computational capabilities that overcome those of traditional classical supercomputers. By processing complex mathematical operations simultaneously rather than sequentially, quantum computers are coming to revolutionize diverse fields, including molecular modeling for drug discovery, advanced logistics optimization, meteorological forecasting, and financial risk modeling [1].

However, this unprecedented computational power introduces several new risks to global cybersecurity frameworks. The principles that enable quantum computers to solve highly complex, multi-variable problems also empower them to solve the mathematical foundations of contemporary cryptography. Classical cryptographic algorithms, which protect from global financial transactions and state secrets to critical industrial operations and citizen privacy, rely on the extreme difficulty of certain mathematical problems, such as prime factorisation or discrete logarithms. A sufficiently powerful quantum computer will solve these problems in a matter of hours or minutes, rendering traditional encryption techniques obsolete [2].

The most important threat arising from the development of quantum computing lies in its ability to compromise asymmetric (public-key) cryptography. Widely deployed protocols such as RSA (Rivest–Shamir–Adleman), Diffie-Hellman, and Elliptic Curve

Cryptography (ECC) form the backbone of modern digital trust. They secure Transport Layer Security (TLS/SSL) for web traffic, virtual private networks (VPNs), digital signatures, and identity authentication protocols across the globe [3].

In 1994, mathematician Peter Shor published Shor’s Algorithm, a quantum algorithm capable of finding the prime factors of an integer in polynomial time. On a classical computer, factoring a large number into its prime components requires exponential time, an operational barrier that guarantees safety. Shor’s algorithm bypasses this barrier, meaning that any public-key infrastructure utilizing RSA or ECC can be systematically decrypted once a Quantum Computer becomes operational [4].

While symmetric cryptographic standards (such as AES-256) are less vulnerable, requiring only an increase in key sizes to maintain security via Grover’s Algorithm, the collapse of public-key cryptography would trigger a catastrophic failure of data confidentiality, systemic integrity, and non-repudiation worldwide.

A critical dimension of this threat is the "Harvest Now, Decrypt Later" strategy currently executed by adversarial nation-states and sophisticated cyber-criminal organizations [5]. Malicious actors are actively intercepting and archiving massive volumes of highly sensitive, encrypted European data, including military communications, state intelligence, intellectual property, financial ledgers, and critical infrastructure blueprints. Even though these actors cannot decrypt the stolen data today, they store it in vast repositories, waiting for the arrival of a Quantum Computing to decrypt it later. Consequently, the quantum threat is not a distant, future concern; it is an active, present-day vulnerability affecting data with long-term security classifications.

### a. The Challenge of Digital Sovereignty

For the European Union, the quantum challenge is directly connected with the broader geopolitical need of digital sovereignty. In an era marked by shifting global alliances, hybrid warfare, and industrial espionage, Europe cannot afford to rely on foreign proprietary technologies to secure its communications. If the core cryptographic components used across EU Member States are manufactured, controlled, or standardized by external entities, Europe compromises its strategic autonomy [6]. Moreover, several vulnerabilities of European critical infrastructure present a severe systemic risk. EU modern digital systems rely on tightly interconnected networks:

- **Smart Electrical Grids and Energy Networks:** Rely on synchronized communication and automated distribution systems. A cryptographic breach could allow adversaries to hijack control systems, inducing widespread power blackouts and industrial disruption.
- **Telecommunications and Data Centres:** Form the main system of the digital economy. The compromise of core transport networks would paralyze financial markets, cloud computing services, and administrative governance.
- **Healthcare Systems:** Manage vast arrays of highly sensitive personal medical records and operate connected hospital equipment. Intercepting or altering this data would compromise patient safety and violate foundational privacy rights.

- **Governmental and Military Communication:** Require absolute, uncompromised confidentiality. Leakage of classified diplomatic or defense telemetry directly threatens European territorial integrity and collective security.

Updating these sometimes legacy systems to quantum-safe is a big challenge. It demands a comprehensive recording of hardware and software tools and protocols, a process that is time demanding and costly.

## 2. The European Approach: A Strategy for Quantum Security

Recognizing the scale and immediacy of this technological challenge, the European Union has issued a comprehensive, proactive strategy to safeguard its digital future [7]. The European approach does not rely on a single defensive mechanism but is based instead on a robust, roadmap combining two core technological areas:

- **Post-Quantum Cryptography (PQC):** This software-driven layer focuses on developing and deploying new mathematical algorithms that are believed to be resistant to both classical and quantum computational attacks. PQC can be integrated into existing network architectures via software updates [8], providing an accessible first line of defense for general consumer applications, commercial enterprises, and standard web traffic [9].
- **Quantum Key Distribution (QKD):** This hardware-driven, physics-based layer provides an unhackable mechanism for secure key exchange. Unlike PQC, which relies on the assumed computational complexity of new mathematical problems, QKD relies on the immutable laws of quantum physics. By transmitting cryptographic keys using the quantum states of individual photons, QKD guarantees that any attempt to intercept, measure, or clone the key alters its physical properties, instantly alerting the communicating parties to the presence of an eavesdropper.

By synthesizing PQC and QKD, the European Union aims to establish a multi-layered, resilient security architecture. PQC provides scalable, algorithmic protection across broad software ecosystems, while QKD offers a mathematically secure shield for the transmission of highly sensitive data across critical state, military, and industrial networks.

### a. EuroQCI: The Blueprint for a Quantum-Secure Europe

The European Quantum Communication Infrastructure (EuroQCI) initiative represents a collaborative attempt aimed at building a highly secure quantum communication network covering the entire European Union. Launched in 2019, the initiative gained momentum as all 27 EU Member States, alongside the European Commission and the European Space Agency (ESA), officially signed the EuroQCI Declaration. This political commitment mandates the creation of a continent-wide infrastructure dedicated to securing Europe's public services, critical networks, and sensitive data structures against quantum and classical cyber threats [10].

The strategic objectives of EuroQCI are integrated into Europe’s Digital Decade 2030 targets [11]. These targets state that the European Union must position itself at the cutting edge of global quantum capabilities by the end of the decade, raining domestic industrial competitiveness, scientific excellence, and complete technological sovereignty. The primary objective of EuroQCI is to transition quantum communication from an isolated, experimental laboratory phenomenon into an operational, reliable, and high-capacity infrastructure that shields the European single market and its governance structures from external disruptions.

To achieve continent-wide coverage, the EuroQCI architecture is divided into two interconnected operational segments: the Terrestrial Segment and the Space Segment.

**The Terrestrial Segment**

The terrestrial segment relies on standard fiber-optic communication networks equipped with dedicated QKD hardware. This segment is primarily responsible for securing communications over short to medium distances (typically up to 100 kilometers per individual link, due to physical photon attenuation within optical fibers). At the local level, this segment is implemented as National Quantum Communication Infrastructures (NatQCIs), which each EU Member State actively deploying it. These national networks connect high-density administrative, financial, and industrial centers via specialized Metropolitan Area Networks (MANs) and regional trunks. To overcome fiber distance limitations, the terrestrial segment employs trusted relay nodes, where quantum keys are decrypted and re-encrypted within heavily secured physical environments, allowing keys to traverse entire nations.

**The Space Segment**

Fiber networks cover urban and national areas, but bridging the big geographic distances required for continent-wide quantum communication requires using an excessive number of trusted nodes, which increases cost and physical security risks. To resolve this, EuroQCI incorporates a comprehensive Space Segment. The Space Segment uses satellites equipped with quantum payload systems capable of generating and transmitting entangled or single photons through the vacuum of space, where photon absorption is minimal. The transmissions occur at Optical Ground Stations (OGSs), which are specialised telescopes that communicate with satellites via lasers rather than radio waves. All OGSs are interconnected with the terrestrial segment of their Member State, for retrieving and relaying cryptographic keys in the wider network.  This allows for long-distance QKD, interconnecting distant Member States, islands, and overseas territories. The primary operational mechanism relies on Low Earth Orbit (LEO) satellite configurations, moving towards highly secure constellations. Key initiatives driving this segment include:

- **The EAGLE-1 Mission:** EAGLE-1 serves as the initial, operational platform to test, validate, and deploy space-based QKD capabilities across Europe, providing the crucial orbital infrastructure needed to feed quantum keys directly into terrestrial networks via OGSs [12].

- **The IRIS² Constellation:** Infrastructure for Resilience, Interconnectivity and Security by Satellite (IRIS²) is Europe’s multi-orbital satellite constellation designed to provide secure, high-speed connectivity for government and civilian applications. IRIS² is designed to integrate quantum communication components, embedding EuroQCI's space-based requirements directly into the core of Europe's next-generation communication infrastructure [13].

## b. Digital Europe Programme and Connecting Europe Facility

The implementation of EuroQCI follows a phased, multi-year framework backed by significant European funding instruments:

### Phase 1: National Foundations (Digital Europe Programme)

The initial phase focused on building national capabilities, fostering domestic supply chains, and establishing operational pilot networks within each individual Member State. Funded primarily through the Digital Europe Programme (DEP), this phase resulted in the creation of 27 distinct NatQCI projects (such as HellasQCI in Greece, CYQCI in Cyprus, BGQCI in Bulgaria, and NL-QCI in the Netherlands). These projects successfully demonstrated the feasibility of QKD within localized contexts, validated multi-vendor hardware interoperability, trained thousands of professionals, and engaged national stakeholders across government, academia, and industry.

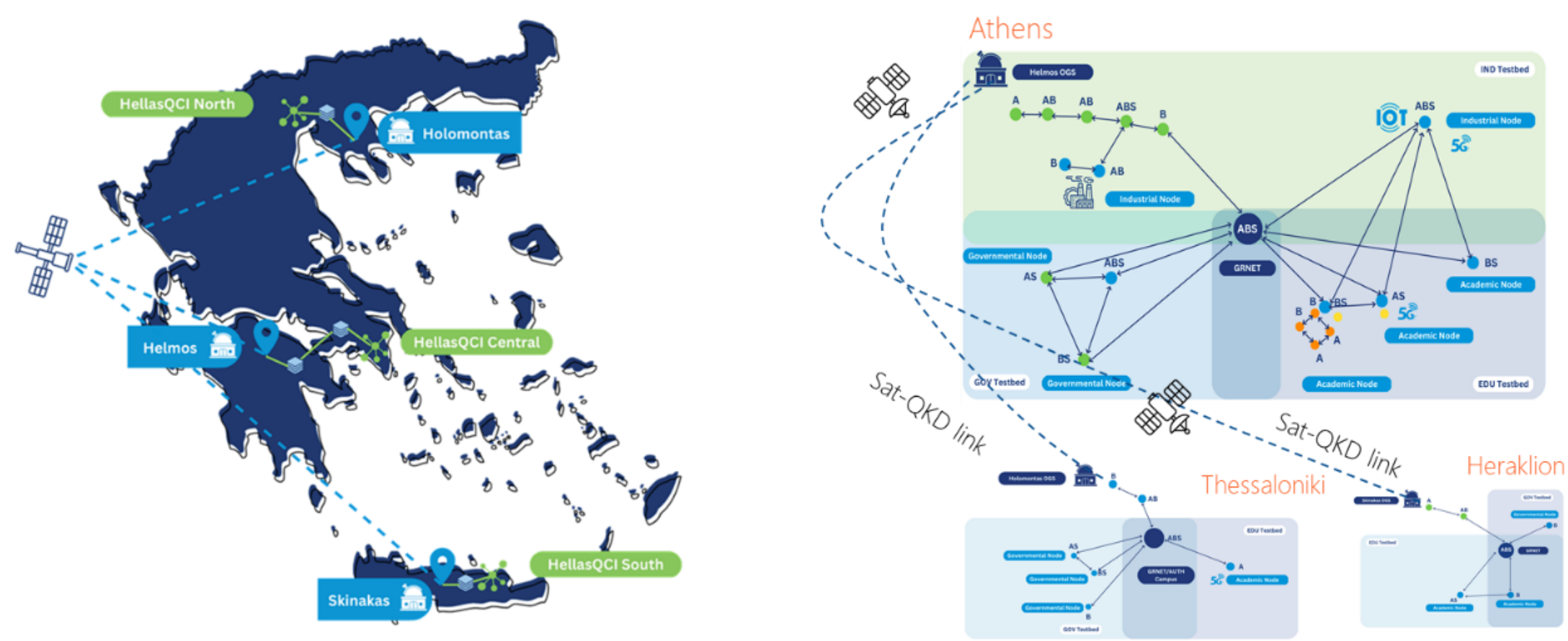

*Figure 1: HellasQCI - The Greek national Quantum Communications Infrastructure, comprising of three terrestrial MAN sites (650km fiber network) in Athens, Thessaloniki and Heraklion (Crete) and three OGSs in Helmos, Holomontas and Skinakas*

### Phase 2: Cross-Border Integration (Connecting Europe Facility - CEF Digital)

With national foundations securely laid, EuroQCI entered its second major phase, funded under the Connecting Europe Facility (CEF) Digital program. This phase marks the transition from isolated, national experiments to an integrated, cross-border network architecture. The primary goal of CEF Digital quantum projects is to interconnect the individual NatQCIs, creating physical and operational corridors that allow secure quantum data flow across national boundaries. By deploying cross-border terrestrial fiber links and upgrading Optical Ground Stations to interface with orbital systems like EAGLE-1, Phase 2 is connecting individual national networks into a singular, continent-wide quantum shield.

## 3. SEEWQCI: Building Europe's South-Eastern to Western Quantum Corridor

The South-East Europe to Western Europe Quantum Communication Infrastructure (SEEWQCI) project stands as one of the most strategically significant initiatives approved under the second phase of the EuroQCI CEF Digital framework [14]. Coordinated by the National Infrastructures for Research and Technology (GRNET) under the auspices of the Greek Ministry of Digital Governance, SEEWQCI forms a cross-border alliance interconnecting four European nations: Greece, Cyprus, Bulgaria, and the Netherlands. The project represents a geostrategic approach to European security, building a secure quantum corridor that links the outer maritime and land boundaries of South-Eastern Europe directly to the highly developed digital hubs of Western Europe.

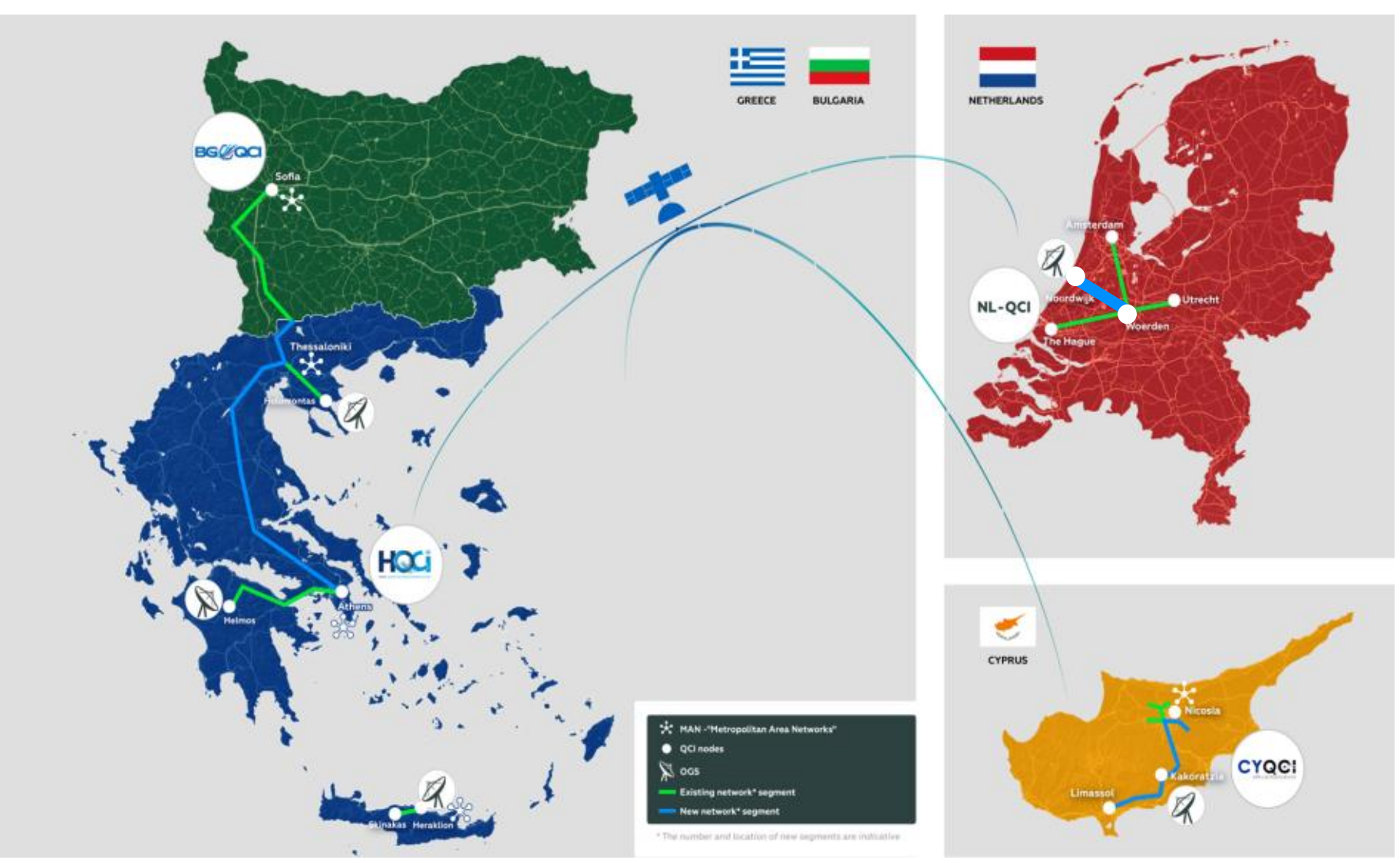

*Figure 2: SEEWQCI foresees the implementation of the Balkan QKD corridor connecting the Helmos OGS to Sofia via a cross-border terrestrial interconnection, and satelite connections between GR-CY-NL-BG*

The consortium includes **5 National Ministries, 5 National Security Authorities, 4 Security Operation Centres (SOCs), 9 Research and Technology Organisations (RTOs) and 3 Industrial Partners.** The SEEWQCI project has six highly defined, strategic objectives aiming to build an operational, resilient, and multi-layered quantum ecosystem:

**Objective 1: The South-East European Terrestrial QKD Corridor**

The project is deploying a 1,100-kilometer cross-border terrestrial QKD fiber backbone connecting Greece and Bulgaria. The terrestrial backbone establishes a concrete foundation for future physical expansions northward and westward, eventually linking the Balkan area directly into Central and Western European quantum hubs. This route provides crucial redundancy and network resilience for the wider European single market.

**Objective 2: Satellite & Hybrid Quantum Connectivity**

To overcome geographical fragmentation, such as the Mediterranean sea separating Cyprus and the Greek islands from mainland Europe, SEEWQCI is establishing advanced space-based connectivity. The project integrates five highly sophisticated Optical Ground Stations (OGSs) situated at strategic astronomical observatories:

- **Greece:** Helmos Observatory (Peloponnese), Holomontas Observatory (Thessaloniki), and Skinakas Observatory (Crete).
- **Cyprus:** Kakoratzia OGS.
- **The Netherlands:** Noordwijk OGS.

These five OGS installations are engineered or upgraded to interface directly with the EAGLE-1 satellite system, creating a total of six distinct cross-border quantum links. This architecture enables:

- **Pure Space Cross-Border Connections:** Establishing direct, satellite-mediated key distribution between Greece, the Netherlands, and Cyprus
- **Hybrid Cross-Border Connections:** Interconnecting distant states via combined terrestrial-space routing, specifically allowing safe data transit.

**Objective 3: Interoperable Key Management & Network Orchestration**

A critical bottleneck in multi-vendor QKD deployments is the coordination of cryptographic keys across different hardware manufacturers and software architectures. Objective 3 addresses this by deploying a unified Key Management System (KMS) and network orchestration framework. This layer abstracts the underlying physical QKD hardware, ensuring that keys generated by different vendors can be seamlessly synchronized, managed, and routed across distinct national domains. Interoperability serves as the operational foundation required to scale cross-border quantum services.

**Objective 4: Real-World Cross-Border Demonstrations**

To validate the reliability, latency, and security of the integrated network, SEEWQCI is executing 29 comprehensive cross-border communication use cases across more than 30 trusted nodes. These large-scale demonstrations involve real end-users, including ministries, national security authorities, and critical infrastructure operators.

**Objective 5: Pan-European Cooperation and Alignment**

SEEWQCI does not operate in isolation; it maintains direct coordination with the broader European quantum ecosystem. The project establishes formal Letters of Support (LoS) and operational links with 10 parallel EuroQCI CEF projects and supports direct collaboration across 25 EU Member States.

**Objective 6: EU Security and Standardisation Alignment**

As quantum hardware matures, it must be accompanied by rigorous certification and standardization frameworks to gain institutional trust. Objective 6 ensures that all SEEWQCI network topologies, encryption devices, and operational protocols align fully with emerging EU governance, security, and standardization criteria.

## a. From National Pilots to a Unified Cross-Border Infrastructure

The implementation of SEEWQCI marks a major evolution from the localized achievements of the Phase 1 National Quantum Networks (NatQCIs) to a unified cross-border infrastructure. Each participating nation brings an existing network foundation that SEEWQCI actively expands and integrates as shown in Table 1:

| **Country** | **Phase 1 Foundation (NatQCI Pilot)** | **Phase 2 Expansion & Integration (SEEWQCI Lifecycle)** |
|---|---|---|
| **Greece (HellasQCI)** | 650 km terrestrial QKD fiber network connecting three urban Metropolitan Area Networks (MANs): Athens, Thessaloniki, and Heraklion-Crete. 16 permanent nodes and 3 localized OGS links. | Expansion to 1,450 km of terrestrial fiber. Direct physical integration with the Bulgarian network and technical upgrade of Helmos, Skinakas, and Holomontas OGSs to connect to the EAGLE-1 satellite system. |
| **Bulgaria (BGQCI)** | 285 km localized terrestrial QKD fiber network serving municipal and administrative centers. | Deployment of a direct, high-capacity terrestrial cross-border QKD link into Greece, unlocking hybrid space-enabled routes to Cyprus and the Netherlands via the Greek infrastructure. |
| **Cyprus (CYQCI)** | Localized terrestrial QKD fiber ring interconnecting national administrative and research institutions. | Extension of physical fiber pathways to peripheral national end-nodes and a complete technical upgrade of the Kakoratzia OGS to establish full orbital connectivity with EAGLE-1. |
| **The Netherlands (NL-QCI)** | Developed terrestrial QKD fiber infrastructure linking administrative hubs within the Randstad region. | Physical and logical connection of the Noordwijk Optical Ground Station to the core NL-QCI network, enabling direct satellite links to Southern Europe via EAGLE-1. |

Table 1. National QCI projects participating in SEEWQCI

Through this structural evolution, SEEWQCI transforms isolated solutions of quantum technology into an integrated, cross-border infrastructure. It exploits years of national investment to prove that quantum cryptography can operate reliably across thousands of kilometers, diverse geological terrains, and multi-vendor technical environments.

# 4. Conclusions

The rapid development of quantum computing represents a dual-edged sword of unprecedented proportions. While its computational rewards promise to redefine human scientific achievement, its capacity to solve modern cryptographic systems poses an existential threat to global security, financial stability, and national sovereignty. The "Harvest Now, Decrypt Later" strategies pursued by global adversaries underscore that the timeline to act is limited and the security of tomorrow's infrastructure must be designed, funded, and deployed today.

Facing with this challenge, the European Union's deployment of the European Quantum Communication Infrastructure (EuroQCI) is crucial. Europe is building a virtually unhackable shield to protect its most critical assets. The dual integration of a Terrestrial Segment and a Space Segment guarantees that secure quantum keys can cover high-density cities, cross vast land borders, and reach remote island territories.

The SEEWQCI project serves as a model for regional cooperation. SEEWQCI demonstrates how nations with diverse geographies and digital profiles can build highly secure infrastructure. It successfully addresses real-world complexities of multi-vendor interoperability, key management orchestration, and EU security standards, proving that a unified quantum shield is achievable. By securing the communication links that connect governmental authorities, military, energy grids, and public healthcare, EuroQCI and its initiatives like SEEWQCI ensure that Europe remains a resilient and technologically sovereign area.

**Acknowledgement:** This work was supported by the project SEEWQCI co-funded by the European Union under the Connecting Europe Facility (CEF) programme, Grant Agreement No 101249531